\documentclass[onecolumn,showpacs,11pt,nofootinbib]{revtex4-2}

\usepackage{axodraw}           
\usepackage{graphicx}          
\usepackage{dcolumn}
\usepackage{bm}
\usepackage{color}
\usepackage{eurosym}
\usepackage{multirow}
\usepackage{amsfonts}
\usepackage{amsmath}
\usepackage{amssymb}
\usepackage[english]{babel}
\usepackage{subcaption}        
\usepackage{hyperref}          
\newcommand{\hs}{\hspace*{0.2cm}}
\newcommand{\bea}{\begin{eqnarray}}
\newcommand{\eea}{\end{eqnarray}}
\newcommand{\crn}{\nonumber \\}

\newcommand{\+}{\dagger}
\newcommand{\La}{\Lambda}
\newcommand{\Vien}[1]{{\color{red}#1}}

\begin{document}

\title{Neutrino phenomenology in a Standard Model extension\\
 with $\mathbf{T^\prime\times Z_{10} \times Z_2}$ symmetry}

 \author{V. V. Vien$^{a}$}
\email{vvvien@ttn.edu.vn}
\affiliation{$^{a}$Department of Physics, Tay Nguyen University, 
Vietnam.}

\author{T. Phong Nguyen$^{b}$}
\email{thanhphong@ctu.edu.vn}
\affiliation{$^{a}$Department of Physics, Can Tho University, 
Vietnam.}
\author{T. D. Tham$^{c}$}
\email{tdtham@pdu.edu.vn (corresponding author)}
\affiliation{$^{c}$Faculty of Natural Science Teachers' Education, Pham Van Dong University, 509 Phan Dinh Phung Street, Cam Thanh Ward, Quang Ngai Province, Vietnam.}

\begin{abstract}
We construct a Standard Model (SM) extension with $T^\prime\times Z_{10} \times Z_2$ symmetry for generating the expected neutrino mass matrix with the relation $(M_\nu)_{13}=(M_\nu)_{31}=-\frac{1}{2}(M_\nu)_{22}$ via the contributions of the Type-I seesaw and Weinberg-type operators. The proposed model possesses viable parameters capable of predicting the neutrino oscillation parameters being in good agreement with recent constraints.
Our analysis reveals the predicted regions for the physical quantities,  given as follows. The two mass squared splittings are $\delta m^2\in (69.360, 79.220)\, \mathrm{meV}^2$ and $\Delta m^2\in (2.484, 2.490)10^3\,\mathrm{meV}^2$ for normal ordering (NO) while $\delta m^2\in (69.450, 79.160)\, \mathrm{meV}^2$ and $\Delta m^2\in (-2.464, -2.456)10^3\,\mathrm{meV}^2$ for inverted ordering (IO).
The lightest neutrino mass is $m_{\ell}\in (36.720, 36.780)$ meV for NO and $m_{\ell}\in (62.220,\, 62.310)$ meV for IO. The sum of neutrino mass is $\sum m_\nu \in (136.700,\, 136.800)$ meV for NO and $\sum m_\nu \in (221.400,\, 221.600)$ meV for IO. Two Majorana phases are predicted to be $\alpha\in (6.367, 6.380)^\circ$ and $\beta\in (6.936, 6.946)^\circ$ for NO while $\alpha \simeq 358.800^\circ$ and $\beta \simeq 0.600^\circ$ for IO. Finally, the effective neutrino mass is $m_{\mathrm{ee}}\in (36.940, 36.980)$ meV for NO and $m_{\mathrm{ee}}\in (76.290, 76.360)$ meV for IO. Based on these results, the Yukawa-like couplings are estimated, which can  naturally explain the charged - lepton as well as neutrino mass hierarchies.

\end{abstract}

\maketitle

\section{Introduction \label{section 1}}
The neutrino mass matrix $(M_\nu)$ structure 
plays an important role for understanding the neutrino phenomenology. Specific forms of neutrino mass matrix can arise from the combination of a given fundamental symmetry with the 
SM or its extensions, as has been investigated in various studies, such as the neutrino minimal SM \cite{seesaw1,nuMSM1,nuMSM2,nuMSM3,nuMSM4,nuMSM5}, texture zeros \,\cite{Low2004,Low2005,Ludl2014,Borgohain2021}, $B-L$ model \cite{VLA2020,VienA42020,VienLong2020}, 
$\mu$-$\tau$ (reflection) symmetry \cite{Harrison2002,Xingrpp2016,Nishi2018,Chamoun2021,Duarah2021,Zhaoepjc2022,Diana2023,Xingrpp2023,VienZ242plb,VienT7plb,Vienz2z4z2cjp}, $\mu$-$\tau$ mixed symmetry \cite{Dey2023,Chakraborty2023}, 
nHDMs \cite{THDM,Antonioplb2020cu,AME07,Kubo13,Arroyo21,WU19,Carcamo2024}, 
331 models \cite{331s,331s1,331s2,331s3,331s4,331s5}, etc. Non-Abelian discrete symmetries have emerged as a superior approach in the explanation of the observed lepton mixing pattern (see, 
Refs. \cite{King2013eh,Altarelli2010gt} and the references therein). At present, the Dirac or Majorana nature of the neutrino is one of the key open problems. Inspired of 
 the results presented in Refs. \cite{Agostini19,Abe2023,Shimizu2024}, we consider the Majorana nature of neutrino. 

The fact that among neutrino oscillation parameters, two neutrino mass-squared splitting $\Delta m^2_{21}=m^2_{2}-m^2_{1}$, $|\Delta m^2_{31}|=|m^2_{3}-m^2_{1}|$, the solar mixing angle $s^2_{12}$ and the reactor mixing angle $s^2_{13}$ have been determined with high accuracy whereas 
the atmospheric mixing angle $s^2_{23}$ has also been determined with reasonably good precision and lies close to its maximal value,
$M_\nu$ can be constructed from another approach 
whereby  the phenomenological outcomes stem from applying certain conditions on the entries of $M_\nu$ 
\cite{Dey2023,Chakraborty2023,Chakraborty2024ess,Chakraborty24,VienEPJPlus25}. In recent studies \cite{Chakraborty24,VienEPJPlus25}, a simple relationship between entries of $M_\nu$, 
 $(M_\nu)_{13}=(M_\nu)_{31}=-\frac{1}{2}(M_\nu)_{22}$, has been 
suggested and $M_\nu$ 
takes the form
\begin{equation}
M_\nu=\left(
\begin{array}{ccc}
A & B & F\\
B & -2F & G\\
F & G & J
\end{array}
\right),
\label{Mnu}
\end{equation}
in which,
\begin{itemize}
  \item [$\bullet$] in Ref. \cite{Chakraborty24}, $A, B, F, G$ and $J$ are complex, 
  and 
  the SM is supplemented by $A_4 \times Z_{10} \times Z_2$ symmetry with 
  two $SU(2)_L$ triplets, one 
  doublet and up to thirteen singlets are introduced to generate the desired form of $M_\nu$. 
  \item [$\bullet$] in 
  Ref. \cite{VienEPJPlus25}, $A, B, F$ and $G$ are complex 
  while $J$ is considered as a real parameter. On the other hand, the 
  symmetry of the SM is supplemented by $A_4 \times Z_{12} \times Z_2\times Z_3$ symmetry with one $SU(2)_L$ doublet and eleven 
  singlets are introduced.
\end{itemize}
Based on the above analysis, it is necessary to develop a more optimized model with fewer scalars. In this work, we consider an alternative symmetry, $T^\prime$, featuring a reduced scalar sector compared to Ref. \cite{Chakraborty24}, and the additional Abelian symmetries are fewer than those in Ref. \cite{VienEPJPlus25}. Namely, two suplemented symmetries, $Z_{10}$ and $Z_{2}$, are introduced, and the model involves twelve scalars, including one doublet and eleven singlets. It is noted that $T^\prime$ is the double covering group of $A_4$; however, they are two completely different groups. $A_4$ includes two groups $Z_3$ and $Z_2\times Z_2$ as its subgroups, whereas $T^\prime$ includes three groups $Z_4, Z_6$ and $Q_4$ as its subgroups \cite{Kobayashi2022}.

The structure of the work is outlined as follows. In Section \ref{diagonalization}, we provide a brief presentation of the diagonalization of the Majorana neutrino mass matrix. Section \ref{model} is devoted to the description of the model. 
Section \ref{Numericalanalysis} is dedicated for the numerical analysis. 
Finally, the conclusions are 
presented in Section \ref{conclusion}. Appendices \ref{apptprime} and \ref{appz10} give brief descriptions of $T^\prime$ and $Z_{10}$ groups, respectively. All terms, up to dimension six, 
prevented by the symmetries of the model are summarized in Appendix \ref{preventedterm}. A concise description of the minimization condition of the scalar potential is presented in Appendix \ref{potential}.

\section{\label{diagonalization}Diagonalization of neutrino mass matrix $M_\nu$}
The lepton mixing matrix $U$ is defined as $U= U^{\dagger}_l U_{\nu}$ 
with $U_{\nu}$ and $U_l$ respectively diagonalize the neutrino ($M_\nu$) and charged-lepton ($M_l$) mass matrices. The matrix
$U$ 
is parameterized as follows
\begin{equation}
U= P_l. U_0. P_\nu, \label{U}
\end{equation}
where $P_\nu=\mathrm{diag}(e^{i \alpha}, e^{i \beta},1)$ with $\alpha$ and $\beta$ are two Majorana phases, $P_l$ 
depends on three un-physical phases which can be eliminated 
by phase redefinition of the charged-lepton fields \cite{XingZhou2010},  and PMNS matrix $U_0$ 
is parameterized as \cite{PDG2024},
\bea
U_0  = U^{}_{23} U^{}_{13} U^{}_{12}, \label{U0}
\eea
with
\begin{equation}
U_{12}=
\begin{pmatrix}
c^{}_{12} & s^{}_{12} & 0 \\
-s^{}_{12}& c^{}_{12}& 0 \\
0&0& 1 \end{pmatrix}, \hs
U_{13}=
\begin{pmatrix}
c^{}_{13} & 0 & s_{13}.e^{-i\delta} \\
0&1 &0 \\
-s_{13} .e^{i\delta} & 0 & c^{}_{13} \end{pmatrix},
\hs U^{}_{23}=\begin{pmatrix}
1 & 0 & 0 \\
0 & c_{23} &s^{}_{23} \\
0 & - s_{23}& c^{}_{23} \end{pmatrix}. \label{Uij}
\end{equation}
In the flavor basis (where 
$M_l$ is diagonal), 
the matrix $U$ is that of neutrinos. In this case three active neutrino masses ($m^{}_1$,
$m^{}_2$, $m^{}_3$), three neutrino mixing angles ($\theta^{}_{12}$,
$\theta^{}_{13}$, $\theta^{}_{23}$) and 
CP-violating phases
($\rho$, $\sigma$, $\delta$) can be obtained by diagonalizing  $M^{}_\nu$. 

In the case where un-physical phases 
have been eliminated, the 
matrix $U$ in Eq. (\ref{U}) 
becomes $U=U_0 P_\nu,$ and $M_\nu$ in 
(\ref{Mnu}) 
is diagonalized in the form, $U^\dagger M_\nu U^* 
=\mathrm{diag}(m_1,\,\,m_2,\,\,m_3)$, which implies
\begin{eqnarray}
&&\begin{pmatrix}
m_{11} & m_{12} & m_{13} \\
m_{12} & m_{22} & m_{23} \\
m_{13} & m_{23} & m_{33} \\
 \end{pmatrix} \equiv U_0^\dagger M_\nu U^*_0 =  
\begin{pmatrix}
e^{2i\alpha} m_1 & 0 & 0 \\ 0 & e^{2i\beta}m_2 & 0 \\ 0 &
0 & m^{}_3 \end{pmatrix}. \label{identity}
\end{eqnarray}

Expression (\ref{identity}) leads to following conditions:
\begin{eqnarray}
&&\mathrm{Im} m_{12}=0,\hs \mathrm{Re} m_{12}=0, \hs  \mathrm{Im} m_{13}=0,\crn
&& \mathrm{Re} m_{13}=0,  \hs 
\mathrm{Im} m_{23}=0,\hs \mathrm{Re} m_{23}=0,\hs \mathrm{Im} m_{33}=0. \label{condition}
\end{eqnarray}

Considering the case where $P_\nu =\mathrm{diag}\left(e^{i\alpha},\, e^{i\beta},\, 1\right)$, 
the 
"33" entry of the matrix in Eq. (\ref{identity}) is real \cite{VienEPJPlus25}.
Two Majorana phases $\beta$ and $\alpha$ 
are determined by, 
\begin{eqnarray}
&&\beta= 
\frac{1}{2} \arctan\left(\frac{\mathrm{Im} m_{22}}{\mathrm{Re} m_{22}}\right), \hs \alpha=
\frac{1}{2} \arctan\left(\frac{\mathrm{Im} m_{11}}{\mathrm{Re} m_{11}}\right). \label{alphabeta}
\end{eqnarray}
Three active neutrino masses ($m_1, m_2, m_3$), 
their sum ($\sum m_\nu$), and 
two squared mass differences $\Delta m^2_{21}$ 
and $\Delta m^2_{31}$ 
are given by,
\begin{eqnarray}
&&m_1^2=\big(\mathrm{Im} m_{11}\big)^2+\big(\mathrm{Re} m_{11}\big)^2, \hs m_2^2=\big(\mathrm{Im} m_{22}\big)^2+\big(\mathrm{Re} m_{22}\big)^2, \hs m_3^2=\mathrm{Re}m_{33}^2, \label{m123}\\
&&\sum m_\nu=\sqrt{\big(\mathrm{Im} m_{11}\big)^2+\big(\mathrm{Re} m_{11}\big)^2}+\sqrt{\big(\mathrm{Im} m_{22}\big)^2+\big(\mathrm{Re} m_{22}\big)^2}+\mathrm{Re} m_{33}, \label{summi}\\
&&
\delta m^2=\big(\mathrm{Im} m_{22}\big)^2+\big(\mathrm{Re} m_{22}\big)^2-\big(\mathrm{Im} m_{11}\big)^2-\big(\mathrm{Re} m_{11}\big)^2,  \label{Dm21}\\
&&
\Delta m^2=\big(\mathrm{Re} m_{33}\big)^2-\frac{1}{2}\left[\big(\mathrm{Im} m_{11}\big)^2+\big(\mathrm{Re} m_{11}\big)^2+\big(\mathrm{Im} m_{22}\big)^2+\big(\mathrm{Re} m_{22}\big)^2\right]. \label{Dm31}
\end{eqnarray}

For simplicity, 
$J$ is considered as a real parameter. Hence, the matrix $M_\nu$ 
depends on nine free parameters 
 $J$, $\mathrm{Im}\mathbf{X}$ and $\mathrm{Re}\mathbf{X}$ with $\mathbf{X}=\{A, B, F, G\}$. However, due to the constraint equations (\ref{condition}), the number of free parameters is reduced to two, namely $\mathrm{Re}G$ and $J$.
As a result, $\alpha, \beta$, $m_{i} (i=1,2,3)$, $\sum m_\nu$, $\delta m^2$ and $\Delta m^2$  
can be expressed in terms of six parameters $\theta_{12}, \theta_{23}, \theta_{13}$, $s_\delta$, $\mathrm{Re}G$ and $J$.


\section{\label{model}The Model}
In this section, we show that the 
neutrino mass $M_\nu$ in Eq. (\ref{Mnu}) can be 
produced by 
the combination of the SM  with 
$T^\prime$ symmetry \cite{Kobayashi2022} and two 
Abelian discrete symmetries $Z_{10}$ and $Z_2$. Brief descriptions of $T^\prime$ and $Z_{10}$ groups are respectively presented in Appendices
\ref{apptprime} and \ref{appz10}, respectively. The field content of the SM is added by three right-handed neutrinos $\nu_{iR} \, (i=1,2,3)$ and five singlet scalar fields $\phi, \varphi, \chi, \rho$ and $\eta$. The particle content of the 
model 
is summarised in Table \ref{particleTp2025}.

\begin{table*}[ht]
\caption{\label{particleTp2025}Transformation properties of 
fields under $SU(2)_L\times U(1)_Y\times T^\prime\times Z_{10}\times Z_2$.}
\vspace{0.25 cm}
\centering
\begin{tabular}{ccccccccccccccc|c|c|c|c|c|c|c|c|c|c|c|}\hline
   & $\psi_{L}$ & $l_{1 R}$&$l_{2 R}$&$l_{3 R}$ &$\nu_{1R}$&$\nu_{2R}$&$\nu_{3R}$ & $H$ & $\phi$&$\varphi$ & $\chi$& $\rho$ &$\eta$
\\ \hline 
$SU(2)_{L}$ & 2 &1&1&1 & 1 &$1$&$1$& 2 &$1$&$1$& 1 & 1 & 1\\ 
$\mathrm{U}(1)_Y$  & $-\frac{1}{2}$&$-1$&$-1$&$-1$&$0$&$0$&$0$&$\frac{1}{2}$&$0$&$0$&$0$&$0$ &$0$\\ 
$T^\prime$ & $\underline{3}$& $\underline{1}^{\prime\prime}$&$\underline{1}^{}$&$ \underline{1}^{\prime}$
&$\underline{1}^{\prime}$&$\underline{1}^{}$&$\underline{1}^{\prime\prime}$& $\underline{1}^{\prime}$ & $\underline{3}$ &$\underline{1}$& $\underline{3}$ & $\underline{3}$ &$\underline{1}^{\prime}$ \\
$Z_{10}$ &$3$& $7$&$2$&$2$ & $2$&$2$&$2$ &$0$& $1$&$5$&$6$&$6$ &$6$\\ 
$Z_2$ &$+$& $-$&$+$&$+$ &$+$&$+$&$+$ & $+$&$+$&$-$&$+$&$+$ &$+$ \\\hline
\end{tabular}
\vspace{0.25 cm}
\end{table*}

The particle content, Table \ref{particleTp2025}, provides the following invariant Yukawa terms,
\begin{eqnarray}
- \mathcal{L}_Y &=& \frac{h_1}{\Lambda^2} (\overline{\psi}_{L} \phi)_{\underline{1}} (H \varphi l_{1R})_{\underline{1}}
+\frac{h_2}{\Lambda} (\overline{\psi}_{L}\phi)_{\underline{1}^{\prime}} (H  l_{2R})_{\underline{1}^{\prime\prime}}
+ \frac{h_3}{\Lambda} (\overline{\psi}_{L}\phi)_{\underline{1}^{\prime\prime}} (H  l_{3R})_{\underline{1}^{\prime}} \nonumber\\
&+& \frac{x_1}{\Lambda^2}\,(\overline{\psi}_{L}\, \psi_{L}^c)_{\underline{3}_s} (\widetilde{H}^2\chi)_{\underline{3}}
+\frac{x_2}{\Lambda^2}\,(\overline{\psi}_{L}\, \psi_{L}^c)_{\underline{3}_s} (\widetilde{H}^2\rho)_{\underline{3}}
+ \frac{x_3}{\Lambda^2}\,(\overline{\psi}_{L}\, \psi_{L}^c)_{\underline{1}^{\prime}} (\widetilde{H}^2\eta)_{\underline{1}^{\prime\prime}} \nonumber\\
&+&\frac{y_1}{\Lambda}\,(\overline{\psi}_{L} \phi)_{\underline{1}}(\widetilde{H} \nu_{1R})_{\underline{1}}
+\frac{y_2}{\Lambda} (\overline{\psi}_{L} \phi)_{\underline{1}^{\prime}} (\widetilde{H} \nu_{2R})_{\underline{1}^{\prime\prime}}
+ \frac{y_3}{\Lambda}(\overline{\psi}_{L} \phi)_{\underline{1}^{\prime\prime}} (\widetilde{H}\nu_{3R})_{\underline{1}^{\prime}} \nonumber\\
&+& \frac{z_1}{2} (\overline{\nu}^c_{1R}\nu_{1R})_{\underline{1}} \eta  + \frac{z_2}{2} (\overline{\nu}^c_{2R}\nu_{3R}+\overline{\nu}^c_{3R}\nu_{2R})_{\underline{1}} \eta +h.c, \label{Lleptp}
\end{eqnarray}
where $\Lambda$ is the cut-off scale, $\widetilde{H} = i \tau_2 H^*$ (with $\tau_2$ is the second Pauli matrix), $h_{i}, x_{i}, y_{i}$ and $z_{j}\, (i=1,2,3; j=1,2)$ are 
Yukawa-like couplings.

Together with $T^\prime$ symmetry, the additional 
symmetries $Z_{10}$ and $Z_{2}$ play an 
indispensable role in 
suppressing various unwanted terms, leading to the desired forms of lepton mass matrices.
Namely, in the absence of $T^{\prime}$, six 
 terms $(\overline{\nu}^c_{2R} \nu_{2R})\chi, (\overline{\nu}^c_{2R} \nu_{2R})\rho, (\overline{\nu}^c_{2R} \nu_{2R})\eta$,$(\overline{\nu}^c_{3R} \nu_{3R})\chi, (\overline{\nu}^c_{3R} \nu_{3R})\rho$ and $(\overline{\nu}^c_{3R} \nu_{3R})\eta$ will be allowed and
contribute to the entries "22" and "33" of 
$(M_R)$ in Eq. (\ref{MR}). On the other hand, in the absence of $Z_{10}$, the following terms $(\overline{\psi}_{L} l_{2R})_{\underline{3}}(H\phi^*)_{\underline{3}}, (\overline{\psi}_{L} l_{3R})_{\underline{3}}(H\phi^*)_{\underline{3}}$, $(\overline{\psi}_{L} l_{2R})_{\underline{3}}(H\chi)_{\underline{3}}$, $(\overline{\psi}_{L} l_{2R})_{\underline{3}}(H\chi^*)_{\underline{3}}$, $(\overline{\psi}_{L} l_{2R})_{\underline{3}}(H\rho)_{\underline{3}}$, $(\overline{\psi}_{L} l_{2R})_{\underline{3}}(H\rho^*)_{\underline{3}}$, $(\overline{\psi}_{L} l_{3R})_{\underline{3}}(H\chi)_{\underline{3}}$, $(\overline{\psi}_{L} l_{3R})_{\underline{3}}(H\chi^*)_{\underline{3}}$, $ (\overline{\psi}_{L} l_{3R})_{\underline{3}}(H\rho)_{\underline{3}}$ and $ (\overline{\psi}_{L} l_{3R})_{\underline{3}}(H\rho^*)_{\underline{3}}$ will be invariant and 
contribute to the entries "12, 22, 32, 13, 23" and "33" of 
 $M_l$ in Eq. (\ref{ML}), and the term $(\overline{\nu}^c_{3R} \nu_{3R})\eta^*$ will be invariant and contributes to the entry "33" of 
$M_R$. Furthermore, in the absence of $Z_{2}$, two 
terms $(\overline{\psi}_{L} l_{1R})(H\chi)$ and $(\overline{\psi}_{L} l_{1R})(H\rho)$ will be invariant and contribute to the entries "11", "21" and "31" of 
$M_l$, etc.
As consequences, the structures of lepton mass matrices will be changed and they cannot account for the observed data. 

The following are some comments:
\begin{itemize}
    \item [(1)] $(\overline{\psi}_{L} l_{k R})_{\underline{3}}(H\phi^{*2})_{\underline{3}_{a}}=0$ and $(\overline{\nu}^c_{1R} \nu_{3R})_{\underline{1}}(\phi^{2})_{\underline{1}}=0$ due to the antisymmetry of $\underline{3}_a$ as a result of 
        $\underline{3}\times \underline{3}$ of $T^\prime$ and the VEV alignment of $\phi$.
    \item [(2)] $(\overline{\psi}_{L} \psi^c_{L})_{\underline{3}_a}=0$ due to the antisymmetry of $\underline{3}_a$ as a result of 
        $\underline{3}\times \underline{3}$ of $T^\prime$ symmetry.
    \item [(3)] $(\overline{\nu}^c_{1R} \nu_{1R})_{\underline{1}^{\prime}}(\phi^{2})_{\underline{1}^{\prime\prime}}=0$, $(\overline{\nu}^c_{2R} \nu_{3R})_{\underline{1}^{\prime}}(\phi^{2})_{\underline{1}^{\prime\prime}}=0$, $(\overline{\nu}^c_{3R} \nu_{2R})_{\underline{1}^{\prime}}(\phi^{2})_{\underline{1}^{\prime\prime}}=0$,
        $(\overline{\nu}^c_{1R} \nu_{2R})_{\underline{1}^{\prime\prime}}(\phi^{2})_{\underline{1}^{\prime}}=0$, $(\overline{\nu}^c_{2R} \nu_{2R})_{\underline{1}^{\prime\prime}}(\phi^{2})_{\underline{1}^{\prime}}=0$ and $(\overline{\nu}^c_{3R} \nu_{3R})_{\underline{1}^{\prime\prime}}(\phi^{2})_{\underline{1}^{\prime}}=0$ due to the results $\underline{1}^{\prime}$ and $\underline{1}^{\prime\prime}$ as consequences of 
        $\underline{3}\times \underline{3}$ of $T^\prime$ and the VEV alignment of $\phi$.
  \end{itemize}

 The minimization conditions of the scalar potential (see in Appendix \ref{potential})  provide  the following vacuum alignments:
\bea
&&\langle H\rangle =\left(0\hspace{0.35 cm} v\right)^T, \hs\langle\phi\rangle =
 (v_{\phi},\hspace{0.25 cm}0, \hspace{0.25 cm}0),\hs \langle\varphi\rangle =v_\varphi,\hs \langle \eta\rangle=v_\eta, \nonumber\\
&&\langle\chi\rangle =
 (v_\chi, \hspace{0.25 cm} 0, \hspace{0.25 cm} 0), \hspace{0.25 cm}
 \langle\rho\rangle =
  (0,\hspace{0.25 cm}v_\rho, \hspace{0.2 cm} -v_\rho). 
 \label{VEValig}
\eea

The Lagrangian expression in (\ref{Lleptp}) with the VEVs in Eq. (\ref{VEValig}) and the tensor product of $T^\prime$ group in the T-diagonal basis \cite{Kobayashi2022}, after symmetry breaking,
yields the 
structures of charged-lepton mass matrix $(M_{l})$, Dirac neutrino mass matrix $(M_{D})$, left- and right-handed Majorana neutrino mass matrices ($M_{L}, M_{R}$):
\begin{eqnarray}
&&M_{l}= 
\mathrm{diag}\left(h_1 v \lambda_\varphi \lambda_\phi, \hs h_2 v \lambda_\phi, \hs h_3 v \lambda_\phi \right)\equiv \mathrm{diag}\left(m_e, \hs m_\mu, \hs m_\tau \right), \label{Ml}\\
&&M_{L}= 
\left(
\begin{array}{ccc}
-2 b_{L} \hs &c_L-b_{L} & -a_L \\
c_L-b_{L} & 2 a_L \hs& b_{L} \\
-a_L \hs& b_{L} & 2b_{L}+c_L \\
\end{array}%
\right), 
\label{ML} \\
&& M_{D}=
\left(
\begin{array}{ccc}
a_D & 0 & 0 \\
0   & b_D & 0 \\
0   & 0 & c_D
\end{array}%
\right), \hs 
M_{R}=
\left(
\begin{array}{ccc}
a_R \hs & 0 \hs & 0 \\
0 \hs& 0 \hs& b_R \\
0 \hs& b_R \hs& 0 \\
\end{array}%
\right).\label{MR}
\end{eqnarray}
where
\bea
&&\lambda_\varphi=\frac{v_\varphi}{\Lambda}, \hs \lambda_\phi=\frac{v_\phi}{\Lambda},  \label{lavarphiphi}\\
&&a_L=x_1 \lambda^2_H v_\chi, \hs b_L=x_2 \lambda^2_H  v_\rho,\hs c_L=x_3 \lambda^2_H v_\eta, \crn
&&a_D=y_1 \lambda_H v_\phi, \hs b_D=y_2 \lambda_H v_\phi, \hs c_D=y_3 \lambda_H v_\phi, \crn
&& a_R=z_1 v_\eta, \hs b_R=z_2 v_\eta \hs\hs  \left(\lambda=\frac{v}{\Lambda}\right). \label{abclDR}
\eea
In the $(\nu^c_L,\, \nu_R)$ basis, the 
$6\times 6$ neutrino mass
matrix has the following form
\bea M_{6\times 6}=\left(%
\begin{array}{ccc}
M_L & M_{D} \\
M^T_D & M_{R}\\
\end{array}%
\right), \label{Mnu66}\eea
 which can be diagonalized by a matrix $U$  satifying,  
\begin{equation} \label{U66}
U^{\dagger} M_{6\times 6} U^{\ast}=\begin{pmatrix} m_{\nu} & \mathbf{0}\\
\mathbf{0}& m_{N} \end{pmatrix},
\end{equation}
where $m_{\nu}$ and $m_{N}$ are $3\times 3$ diagonal matrices. 
The matrix $U$ is 
parameterised by, 
\begin{align}
&U=\begin{pmatrix} c_\Theta U_{0} & s_\Theta U_{N}^{\ast} \\
-s^\dagger_\Theta U_{0} & c^\dagger_\Theta U_{N}^{\ast} \end{pmatrix},
\hs s_\Theta=\sum_{q=0}^\infty \frac{(-\Theta\Theta^\dagger)^q \Theta}{(2q+1)!}, \hspace{0.25 cm} c_\Theta=\sum_{q=0}^\infty \frac{(-\Theta\Theta^\dagger)^q}{(2q)!},
\label{U66nu}\end{align}
with
 $\Theta$ is the matrix representing the active-heavy neutrino mixing, 
\begin{align}
\Theta = M_D M^{-1}_R=
\left(
\begin{array}{ccc}
 0 & 0 & \frac{a_{D}}{b_{R}} \\
 0 & \frac{b_{D}}{a_{R}} & 0 \\
 \frac{c_{D}}{b_{R}} & 0 & 0 \\
\end{array}
\right), \label{Theta} \end{align}
where $a_D, b_D, c_D, a_R$ and $b_R$ are defined in Eq. (\ref{abclDR}).
Two matrices $U_{0}$ and $U_{N}$ 
are the ones that diagonalize the active ($M_\nu$) and heavy ($M_N$) neutrino mass matrices, respectively,
\begin{align}
&U_{0}^{\dagger} M_{\nu} U_{0}^{\ast}=m_{\nu} \equiv \text{diag}(m_1,m_2,m_3), \label{mlightef}\\
&U^T_{N} M_N U_N=m_{N}\equiv \text{diag}(m_{1N}, m_{2N}, m_{3N})\label{MNdiagDef}.
\end{align}

As shown in Tab. \ref{Heavyparameter1}, in the model under consideration, the active-heavy mixing parameter $\Theta$ is tiny with $|\Theta_{ij}|_{\mathrm{max}} \sim \mathcal{O}\big(10^{-11}\big)\mathbf{I}_{3\times3} \ll \mathbf{I}_{3\times3}$. Consequently, in the first-order approximation in $\Theta$, obtain the following expressions:
\begin{itemize}
\item [$\bullet$] The 
matrix $U$ in (\ref{U66nu}) is reduced to
\begin{eqnarray}
&&U \simeq 
\begin{pmatrix} \left(\mathbf{I}_{3\times3}-\frac{1}{2}\Theta \Theta^{\dagger}\right) U_{0} & \Theta\, U_{N}^{\ast} \\
-\Theta^{\dagger} U_{0} & \left(\mathbf{I}_{3\times3}-\frac{1}{2}\Theta^{\dagger} \Theta\right)U^*_{N}
\end{pmatrix}
\equiv  \begin{pmatrix} V_{\nu} & R_{\nu}\\
S_{\nu} & U_{\nu} \end{pmatrix}
\simeq \begin{pmatrix} U_{0} & \mathbf{0} \\
 \mathbf{0} & U^*_{N} \end{pmatrix}.\label{U66reduced1}
\end{eqnarray}
\item [$\bullet$] Since $|\delta M_{N}|_{\mathrm{max}} =\left|\frac{1}{2}\big(\Theta^{\dagger} \Theta M_R+ M^T_R \Theta^T \Theta^{\ast}\big)\right|_{\mathrm{max}} \sim \mathcal{O}\big(10^{-11}\big) \mathbf{1}_{3\times 3} \big(\mathrm{GeV}\big)\ll |M_R|$, the heavy neutrino mass matrix ($M_{N}$) 
     is approximately equal to 
$M_R$,
\bea
M_{N} = M_R+\frac{1}{2} \Big(\Theta^{\dagger} \Theta M_R + M^T_R \Theta^T \Theta^{\ast}\Big)\simeq \mathrm{M}_R.\label{Mheavyfist}\eea
\item [$\bullet$] The light ($U^{\mathrm{lep}}$) and heavy ($V^{\mathrm{lep}}$) lepton mixing matrices are respectively 
determined as,
\bea
&&U^{\mathrm{lep}}=U^{\dagger}_l\Big(\mathbf{I}_{3\times 3}-\frac{1}{2}\Theta \Theta^{\+}\Big) U_{0}
\simeq U^{\dagger}_l U_{0}\equiv U_{0}, \label{lightmixing}\\
&&V^{\mathrm{lep}}=U^{\dagger}_l\Big(\mathbf{I}_{3\times 3}-\frac{1}{2}\Theta^{\+} \Theta\Big) U_{N}^{\ast} \simeq U^{\dagger}_l U_{N}^{\ast}\equiv U_{N}^{\ast}. \label{heavymixing}
\eea
\item [$\bullet$] The heavy-light  lepton mixing matrices are determined as,
\bea
&&|R^{\mathrm{lep}}|_{\mathrm{max}}=|U^{\dagger}_l \Theta U_{N}^{\ast}|_{\mathrm{max}}\sim \mathcal{O}\big(10^{-11}\big) \mathbf{1}_{3\times 3}\ll  \mathbf{1}_{3\times 3}, \\
&&|S^{\mathrm{lep}}|_{\mathrm{max}}=|- U^{\dagger}_l\Theta^{\dagger} U_{0}|_{\mathrm{max}}\sim \mathcal{O}\big(10^{-11}\big) \mathbf{1}_{3\times 3}\ll  \mathbf{1}_{3\times 3}. \label{lightheavymixing}
\eea
Therefore, the lepton mixing is obtained from Eq. (\ref{U66reduced1}) as follows,
\begin{align}
&&U^{\mathrm{lep}}_{6}=U^{\dagger}_l. U\simeq  \begin{pmatrix} U^{\dagger}_l U_{0} & \mathbf{0}\\
\mathbf{0} & U^{\dagger}_l U_{N}^{\ast} \end{pmatrix}\equiv \begin{pmatrix} U_{0} & \mathbf{0}\\
\mathbf{0} & U_{N}^{\ast} \end{pmatrix}. \label{U66}
\end{align}
\end{itemize}

\subsection{\label{lìhtneutrino}Active neutrino mass}
With the aid of Eqs. (\ref{ML})-(\ref{MR}), the active effective neutrino mass matrix $M_{\nu}$, generated through the combination of type-I and type-II seesaw mechanisms $M_\nu = M_L-M_{D}M^{-1}_{R}M^{T}_{D}$, has the desired structure (\ref{Mnu}), where $A, B, F, G$ and $J$ are complex 
which can be expressed in terms of the model parameters as follows,
\bea
A= \Vien{-}2 b_L-A_d, \hs
B=-b_L + c_L, \hs
F = -a_L, \hs
G = b_L - B_d, \hs
J = 2 b_L + c_L,  \label{ABFGJ}
\eea
where $A_d = \frac{a_D^2}{a_R}, \, B_d = \frac{b_D c_D}{b_R}$ with $a_L, b_L, c_L$, $a_D, b_D, c_D$, $a_{R}$ and $b_R$ are defined in Eqs. (\ref{ML})-(\ref{MR}).
Expression (\ref{ABFGJ}) with the aid of Eq. (\ref{abclDR}) yields the following relations,
\bea
 &&x_1=-3 F \kappa_\chi, \hs x_2=(J - B) \kappa_\rho,\hs  x_3= (J + 2 B)\kappa_\eta, \crn
 &&z_1=\frac{y_1^2}{(2 B-3 A-2 J) \kappa_\phi}\Big(\frac{v_\phi}{v_\eta}\Big), \,\,\,
 \hs z_2=\Big(\frac{3 A-2 B+2 J}{B+3 G-J}\Big) \Big(\frac{y_{2} y_{3}}{y_{1}^2}\Big) z_1,\label{Relation}
\eea
where $\kappa_\gamma 
= \frac{\La^2}{3 v^2 v_\gamma}\,\,(\gamma=\phi, \chi, \rho, \eta)$, 
and $A, B, F$ and $\mathrm{ImG}$ can be expressed in terms of three mixing angles $\theta_{12}, \theta_{23}, \theta_{13}$, Dirac CP phase $\delta$,
$\mathrm{Re} G$ and $J$ by using Eqs. (\ref{Mnu}) and (\ref{U0})-(\ref{condition}). In this sense, $y_1, y_2$ and $y_3$ play the role of input parameters.

The diagonalization of 
$M_\nu$ to determine the light neutrino masses and the corresponding neutrino mixing matrix is performed in Sec. \ref{diagonalization}.
\subsection{\label{heavyneutrino}Heavy neutrino mass}
The heavy neutrino mass matrix, determined in  Eqs. (\ref{MR}) and (\ref{Mheavyfist}), has following eigenvalues and mixing matrix,
\bea && m_{1,3N} 
=|b_R| + \frac{|a_D|^2 + |c_D|^2}{2 |b_R|}, \,\, m_{2N} 
=|a_R| + \frac{|b_D|^2}{|a_R|}, \label{M123}\\
&&U_N=\frac{1}{\sqrt{2}}\left(
\begin{array}{ccc}
 1 & 0 & 1 \\
 0 & \sqrt{2} & 0 \\
 -1 & 0 & 1 \\
\end{array}
\right), \label{UN}\eea
where $a_D, b_D, c_D$, $a_B$ and $b_R$ are defined in Eqs. (\ref{ML})-(\ref{MR}).
\subsection{\label{nu2betadecay}Neutrinoless double beta decay}
The effective neutrino mass related to the neutrinoless double beta ($0\nu 2 \beta$) decay which is determined 
is given as follows \cite{Abada2021mee,Das2024mee}:
\begin{eqnarray}
&&m_{ee} = \left|\sum^3_{i=1} \big[(U^{\mathrm{lep}})_{1 i}\big]^2 m_i + \sum_{I=1}^{3} \big[(R^{\mathrm{lep}})_{1 I}\big]^2 \left(\frac{m_{IN}}{p^2-m_{IN}^2}\right)p^2 \right|, \label{meegeneralz2z4z2}\end{eqnarray}
where $p^2=
-10^4 \mathrm{MeV}^2$ is the virtual momentum exchanged.
As shown in Tabs. \ref{Heavyparameter1} and \ref{Heavyparameter2}, the heavy neutrino masses $m_{IN} \sim \mathcal{O}(10^{5})\,\mathrm{GeV}$ for NO [$m_{IN} \sim \mathcal{O}(10^{4})\,\mathrm{GeV}$ for IO] and
the heavy-light mixings are tiny $(R^{\mathrm{lep}})_{1I} \sim \mathcal{O}\big(10^{-9}\big)$ for NO [$(R^{\mathrm{lep}})_{1I} \sim \mathcal{O}(10^{-8})$ for IO]; thus, the contribution of 
the second term in Eq. (\ref{meegeneralz2z4z2}) 
is negligible, and $m_{ee}$ 
is reduced to 
\begin{eqnarray}
 m_{ee}&\simeq&  \left|\sum^3_{i=1} \big[(U^{\mathrm{lep}})_{1 i}\big]^2 m_i\right|
 =\left\{\left[c_{13}^2 \left(c_{12}^2 m_{1} s_{2\alpha}+m_{2} s_{12}^2 s_{2\beta}\right)
-m_{3} s_{13}^2 s_{2 \delta}\right]^2 \right.\crn
&&\hspace{2.95 cm}+\left.\left[c_{13}^2 \left(c_{12}^2 m_{1} c_{2\alpha}+ m_{2} s_{12}^2 c_{2\beta}\right)
+m_{3} s_{13}^2 c_{2\delta}\right]^2\right\}^{\frac{1}{2}}, \label{meeexpress}
\end{eqnarray}
where $m_{1}, m_2$ and $m_3$ are given in Eq. (\ref{m123}), and the lightest neutrino is $m_{\ell}=m_1$ for NO and $m_{\ell}=m_3$ for IO. Equations (\ref{U0})-(\ref{m123}) and (\ref{meeexpress}) tell us that $m_{ee}$ can be expressed in terms of five parameters $\theta_{12}, \theta_{13}$, $\delta_{CP}, \mathrm{Re}G$ and $J$.

\section{\label{Numericalanalysis}Numerical Analysis}
\subsection{Allowed parameters space of the model \label{Allowedparameter}}

\begin{itemize}
\item [$\bullet$] Since the model parameters $\alpha, \beta$, $m_{1,2,3}$, $\sum m_\nu$, $\delta m^2$ and $\Delta m^2$ explicitly depend on six parameters $\theta_{12}, \theta_{23}, \theta_{13}$, $\delta_{CP}$, $\mathrm{Re}G$ and $J$ in which four parameters ($\theta_{12}, \theta_{23}, \theta_{13}$ and $\delta_{CP}$) are observed ones which will be fixed at the best-fit points taken from Ref. \cite{Capozziprd25},
    \bea
&&\left\{
\begin{array}{l}
s^2_{12} =0.303, \hspace{0.25 cm}  s^2_{13} = 2.230\times 10^{-2}, \hspace{0.25 cm} s^2_{23} =0.473, \hspace{0.25 cm} \delta_{CP}=1.20 \pi \hspace{0.15 cm} \mbox{for NO,} \\
s^2_{12} =0.303, \hspace{0.25 cm} s^2_{13} = 2.230\times 10^{-2}, \hspace{0.25 cm} s^2_{23} =0.545, \hspace{0.25 cm} \delta_{CP}=1.48 \pi \hspace{0.15 cm} \mbox{for IO.}
\end{array}
\right. \label{bestfitpoints}\eea
The allowed ranges of $\mathrm{Re}G$ and $J$ will be determined from the experimental constraints on
$\delta m^2$ and $\Delta m^2$ taken from Ref. \cite{Capozziprd25}. In particular, the allowed parameter space of
 $\mathrm{Re}G$ and $J$ are given as follows:
 \bea
&&\hspace{-0.75 cm}\left\{
\begin{array}{l}
\mathrm{Re}G \in (22.705, 22.74)\, \mathrm{meV}, \hspace{0.35 cm}  J \in (43.622, 43.667)\, \mathrm{meV}  \hspace{0.785 cm} \mbox{for NO,}\ \  \\
\mathrm{Re}G \in (70.800, 70.860)\, \mathrm{meV}, \hspace{0.15 cm}  J \in (-15.282,\, -15.250)\, \mathrm{meV}   \hspace{0.15 cm} \mbox{for IO.}
\end{array}
\right. \label{GJrange}\eea

\item [$\bullet$] Expression (\ref{Relation}) implies that $x_1\propto \frac{\Lambda^2}{v_\chi}, x_2\propto \frac{\Lambda^2}{v_\rho}$, $x_3\propto \frac{\Lambda^2}{v_\eta}$, $z_1\propto \frac{v^2_\phi y^2_1}{\Lambda^2 v_\eta}$ and $z_2\propto \frac{v^2_\phi y_2 y_3}{\Lambda^2 v_\eta}$. Therefore, in order for the Yukawa-like couplings to be of comparable magnitude,
    the VEVs of the singlet scalars and the cut-off scale are expected to have the following magnitude orders,
\bea
&&v=174\, \mathrm{GeV},\hs  \Lambda=10^{10} \, \mathrm{GeV}, \crn
&&v_\phi =10^{8} \, \mathrm{GeV}, \hs v_\varphi =10^{7} \, \mathrm{GeV}, \hs v_\eta=10^{6} \, \mathrm{GeV}, \hs v_\chi\sim v_\rho= 10^{5} \, \mathrm{GeV}. \label{VEVscale}
\eea
Furthermore, in order for the Yukawa-like couplings $z_1$ and $z_2$ to be of the same order of magnitude and comparable to those of $x_{1, 2, 3}$, we find the allowed ranges of $y_1, y_2$ and $y_3$ as follows
\bea
y_1\sim y_2\sim y_3 \in (1.00, \, 1.25) 10^{-3} \hspace{0.15 cm} \mbox{(NO and IO)}. \label{y1y2y3ranges}
\eea
\end{itemize}
Below are some comments on the allowed parameters:
\begin{itemize}
  \item [(1)] For NO, in the case of $\mathrm{Re} G< 22.705$ or $J>43.667$, the resulting value of $\delta m^2$ exceeds its upper bound reported in Ref. \cite{Capozziprd25}, whereas for  $\mathrm{Re} G> 22.740$ or $J<43.622$, it goes below the lower bound provided in the same reference.
  \item [(2)] In the case of $y_1>1.25\times 10^{-3}$, $|z_1|$ is found to exceed $\mathcal{O} (1.0)$ for NO and $\mathcal{O}(0.1)$ for IO which are beyond the expected domains and hence are not be considered.
\end{itemize}

\subsection{Numerical analysis result \label{Numericalanalysisre}}
Firstly, we consider the charged-lepton masses.  Expression (\ref{Ml}) implies that $h_1 \lambda_\varphi : h_2: h_3 =m_e:m_\mu:m_\tau$.
Furthermore, the data in Ref. \cite{PDG2024} indicate that $m_e: m_\mu: m_\tau \simeq 1: 10^{2}: 10^{3}$. Thus, the charged lepton mass hierarchy is natural achieved in the case of Eq. (\ref{VEVscale}), and the Yukawa couplings $h_1, h_2$ and $h_3$ have the same scale as follows,
\bea
&&h_1= 0.294, \hspace{0.375 cm} h_2 = 6.070\times 10^{-2}, \hs h_3=1.020. \label{h1h2h3pench}
\eea
It is important to note that, in addition to $H$ and $\phi$, the scalar field $\varphi$ plays an indispensable role in ensuring the natural hierarchy of the charged-lepton masses.

Next, we consider the neutrino sector. Equations (\ref{condition})-(\ref{Dm31}) and (\ref{meeexpress}), with the aid of Eqs. (\ref{bestfitpoints}) and  (\ref{GJrange}), yield the possible ranges of 
the parameters $\delta m^2$, $\Delta m^2$, $m_{1, 2, 3}$, $\sum m_{\nu}$, $m_{\mathrm{ee}}$, $\alpha$ and
$\beta$ which are shown in 
Table \ref{modelparameterrangedij}. 
\begin{itemize}
\item [$\bullet$] 
The predicted values of $\delta m^2$ and $\Delta m^2$ are $\delta m^2\in (69.360, 79.220)\, \mathrm{meV}^2$ and $\Delta m^2\in (2.484, 2.490)10^3\,\mathrm{meV}^2$ for NO while $\delta m^2\in (69.450, 79.160)\, \mathrm{meV}^2$ and $\Delta m^2\in (-2.464, -2.456)$ $10^3\,\mathrm{meV}^2$  for IO which are consistent with 3$\sigma$ constraints given in Ref. \cite{Capozziprd25}.
\item [$\bullet$] Three active neutrino masses are predicted to be $m_1\in (36.720, 36.780)$ meV, $m_2\in (37.700, 37.800)$ meV and $m_3\in (62.210, 62.290)$ meV for NO while $m_1\in (79.340,\, 79.400)$ meV, $m_2\in (79.790,\, 79.880)$ meV and $m_3\in (62.220,\, 62.310)$ meV for IO. As a result, the total neutrino mass is predicted in the ranges of $(136.700,\, 136.800)$ meV for NO and 
     $(221.400,\, 221.600)$ meV for IO which can be compared with the recent constraints on 
     $\sum m_\nu$ reported in Refs. \cite{Capozziprd25,Shao24}. 

\item [$\bullet$] The effective neutrino mass 
$m_{\mathrm{ee}}$ is predicted 
in the ranges of $m_{\mathrm{ee}}\in (36.940,\, 36.980)$ meV for NO and $m_{\mathrm{ee}}\in (76.290, 76.360)$ meV for IO which are below the upper bound obtained from the global 
analysis reported in Ref. \cite{Capozziprd25} with $m_{\mathrm{ee}} < 86 \,\mathrm{meV}$, the experimental limits given by KamLAND-Zen \cite{Abe2023} with $m_{\mathrm{ee}} < 36-156 \,\mathrm{meV}$ and CUORE \cite{CUORE20} with $m_{\mathrm{ee}} < 75-350 \,\mathrm{meV}$, 
    and may be 
    accessible to the future experiments \cite{JCao2020mev,Agostini23}.
\end{itemize}
\begin{table}[t]
\vspace{-0.25 cm}
\caption{\label{modelparameterrangedij} Possible ranges of 
$\delta m^2, \Delta m^2 (\mathrm{in\,\, meV}^2)$, $m_{1,2, 3}$, $\sum m_\nu$, $m_{\mathrm{ee}}$ (in meV) and $\alpha, \beta$ (in degree)}
\vspace{-0.2 cm}
\begin{center}
\begin{tabular}{l c c c c c c c c c c c c c |c|} \hline
\multirow{2}{0.75cm}{\centering } & \multicolumn{2}{c}{Normal odering} & \multicolumn{2}{c}{Inverted odering} \\
\cline{2-5}
& Minimum & Maximum  & Minimum  & Maximum  \\ \hline
$\delta m^2$&$69.360$&$79.220$&$69.450$&$79.160$ \\
$\Delta m^2 $&\hspace{0.1 cm}$2.484\times 10^{3}$&$2.490\times 10^{3}$&$-2.464\times 10^{3}$&$-2.456\times 10^{3}$ \\
$m_{1} $&$36.720$&$36.780$&$79.340$&$79.400$ \\
$m_{2} $&$37.700$&$37.800$&$79.790$&$79.880$ \\
$m_{3} $&$62.210$&$62.290$&$62.220$&$62.310$ \\
$\sum m_{\nu} $&$136.700$&$136.800$&$221.400$&$221.600$ \\
$m_{\mathrm{ee}}$&$36.940$&$36.980$&$76.290$&$76.360$ \\
$\alpha$&$6.367$&$6.380$&$358.800$&$358.800$ \\
$\beta$&$6.936$&$6.946$&$0.617$&$0.623$ \\  \hline
\end{tabular}
\end{center}
\end{table}

Similarly, we find 
the possible ranges of $\mathrm{Re}m_{11}, \mathrm{Im}m_{11}, \mathrm{Re}m_{22}, \mathrm{Im}m_{22}$, $\mathrm{Re}m_{33}, \mathrm{Re}A, \mathrm{Im}A$, $\mathrm{Re}B$, $\mathrm{Im}B$, $\mathrm{Im}G, \mathrm{Re}F$ and $\mathrm{Im}F$ which are shown in Table \ref{modelparameterrangem11}.
\begin{table}[t]
\vspace{-0.25 cm}
\caption{\label{modelparameterrangem11} Possible ranges of 
$m_{11}, m_{22}$, $m_{33}, \mathrm{A}, \mathrm{B}, \mathrm{G}$ and $\mathrm{F}$ (in meV)}
\vspace{-0.2 cm}
\begin{center}
\begin{tabular}{l c c c c c c c c c c c c c |c|} \hline
\multirow{2}{0.75cm}{\centering } & \multicolumn{2}{c}{Normal odering} & \multicolumn{2}{c}{Inverted odering} \\
\cline{2-5}
& Minimum & Maximum & Minimum & Maximum  \\ \hline
$\mathrm{Re} m_{11}$&$-35.870$&$-35.820$&$-71.330$&$-71.260$ \\ 
$\mathrm{Im} m_{11}$&$-8.113$&$-8.104$&$3.388$&$3.398$ \\ 
$\mathrm{Re} m_{22}$&$36.600$&$36.690$&$-79.870$&$-79.770$ \\ 
$\mathrm{Im} m_{22}$&$9.052$&$9.061$&$-1.735$&$-1.720$ \\ 
$\mathrm{Re} m_{33}$&$62.210$&$62.290$&$62.220$&$62.310$ \\ \hline

$\mathrm{ReA}$&$-13.310$&$-13.110$&$-79.100$&$-79.020$ \\ 
$\mathrm{ImA}$&$-1.524$&$-1.521$&$1.973$&$1.976$ \\ 

$\mathrm{ReB}$&$30.320$&$30.360$&$0.600$&$0.620$ \\ 
$\mathrm{ImB}$&$8.783$&$8.793$&$-3.459$&$-3.446$ \\ 

$\mathrm{ImG}$&$-1.6833$&$-1.680$&$0.177$&$0.181$ \\ 

$\mathrm{ReF}$&$-15.880$&$-15.860$&$0.868$&$0.888$ \\ 
$\mathrm{ImF}$&$-2.070$&$-2.060$&$-4.435\times 10^{-3}$&$-6.781\times 10^{-4}$ \\ 

 \hline
\end{tabular}
\end{center}
\end{table}
Furthermore, equation (\ref{Relation}), with the help of Eqs. (\ref{bestfitpoints})-(\ref{y1y2y3ranges}), provides the possible ranges of 
$|x_1|, |x_2|, |x_3|, |z_1|$ and $|z_2|$  which are 
presented in Table \ref{Yukawacouplingrangexyz}. Finally, the light-heavy mixing parameters and heavy neutrino masses are determined in Tables \ref{Heavyparameter1} and \ref{Heavyparameter2}, respectively.
\begin{table}[t]
\caption{\label{Yukawacouplingrangexyz} 
Possible ranges of Yukawa-like couplings $|x_{1}|, |x_2|, |x_3|, |z_1|$ and $|z_2|$}
\vspace{-0.2 cm}
\begin{center}
\begin{tabular}{l c c c c c c c c c c c c c |c|} \hline
\multirow{2}{0.75cm}{\centering } & \multicolumn{2}{c}{Normal odering} & \multicolumn{2}{c}{Inverted odering} \\
\cline{2-5}
& Minimum  & Maximum & Minimum  & Maximum  \\ \hline
$|x_1|$ &$0.528$&$0.529$   &$2.870\times 10^{-2}$&$2.930\times 10^{-2}$ \\ 

$|x_2|$ &$0.175$&$0.176$   &$0.1787$&$0.179$ \\ 

$|x_3|$ &$0.116$&$0.117$   &$1.720\times  10^{-2}$&$1.730\times 10^{-2}$ \\ 

$|z_1|$ &$0.354$&$0.555$   &$3.372\times  10^{-2}$&$5.266\times 10^{-2}$ \\ 
$|z_2|$ &$0.165$&$0.258$   &$3.978\times  10^{-2}$&$6.212\times 10^{-2}$ \\ \hline

\end{tabular}
\end{center}
\end{table}

\begin{table}[t]
\caption{\label{Heavyparameter1} 
\Vien{P}ossible ranges of $\Theta_{ij}, (\Theta\Theta^{\dagger})_{ij}, R^{\mathrm{lep}}_{ij}$ and $S^{\mathrm{lep}}_{ij}$}
\vspace{-0.2 cm}
\begin{center}
\begin{tabular}{l c c c c c c c c c c c c c |c|} \hline
\multirow{2}{0.75cm}{\centering } & \multicolumn{2}{c}{Normal odering} & \multicolumn{2}{c}{Inverted odering} \\
\cline{2-5}
& Minimum  & Maximum  & Minimum  & Maximum  \\ \hline
$|\Theta_{13}|$ &$9.460\times 10^{-9}$&$9.462\times 10^{-9}$   &$3.926\times 10^{-8}$&$3.929\times 10^{-8}$ \\ 
$|\Theta_{22}|$ &$4.402\times 10^{-9}$&$4.403\times 10^{-9}$   &$4.627\times 10^{-8}$&$4.634\times 10^{-8}$ \\ 
$|\Theta_{31}|$ &$9.461\times 10^{-9}$&$9.461\times 10^{-9}$   &$3.927\times 10^{-8}$&$3.928\times 10^{-8}$ \\  \hline

$|(\Theta \Theta^{\dagger})_{11}|$ \hspace{0.1 cm}&$8.951\times 10^{-17}$&$8.951\times 10^{-17}$   &$1.543\times 10^{-15}$&$1.543\times 10^{-15}$ \\
$|(\Theta \Theta^{\dagger})_{22}|$ \hspace{0.1 cm}&$1.938\times 10^{-17}$&$1.938\times 10^{-17}$   &$2.144\times 10^{-15}$&$2.144\times 10^{-15}$ \\
$|(\Theta \Theta^{\dagger})_{33}|$ \hspace{0.1 cm}&$8.951\times 10^{-17}$&$8.951\times 10^{-17}$   &$1.543\times 10^{-15}$&$1.543\times 10^{-15}$ \\  \hline

$|R^{\mathrm{lep}}_{11}|$ &$6.690\times 10^{-9}$&$6.690\times 10^{-9}$   &$2.777\times 10^{-8}$&$2.778\times 10^{-8}$ \\
$|R^{\mathrm{lep}}_{13}|$ &$6.690\times 10^{-9}$&$6.690\times 10^{-9}$   &$2.777\times 10^{-8}$&$2.778\times 10^{-8}$ \\
$|R^{\mathrm{lep}}_{22}|$ &$4.402\times 10^{-9}$&$4.403\times 10^{-9}$   &$4.627\times 10^{-8}$&$4.634\times 10^{-8}$ \\
$|R^{\mathrm{lep}}_{31}|$ &$6.690\times 10^{-9}$&$6.690\times 10^{-9}$   &$2.777\times 10^{-8}$&$2.778\times 10^{-8}$ \\
$|R^{\mathrm{lep}}_{33}|$ &$6.690\times 10^{-9}$&$6.690\times 10^{-9}$   &$2.777\times 10^{-8}$&$2.778\times 10^{-8}$ \\ \hline

$|S^{\mathrm{lep}}_{11}|$ &$2.932\times 10^{-9}$&$2.933\times 10^{-9}$   &$1.609\times 10^{-8}$&$1.610\times 10^{-8}$ \\
$|S^{\mathrm{lep}}_{12}|$ &$5.898\times 10^{-9}$&$5.899\times 10^{-9}$   &$2.444\times 10^{-8}$&$2.444\times 10^{-8}$ \\
$|S^{\mathrm{lep}}_{13}|$ &$6.791\times 10^{-9}$&$6.791\times 10^{-9}$   &$2.620\times 10^{-8}$&$2.620\times 10^{-8}$ \\
$|S^{\mathrm{lep}}_{21}|$ &$2.076\times 10^{-9}$&$2.077\times 10^{-9}$   &$1.796\times 10^{-8}$&$1.798\times 10^{-8}$ \\
$|S^{\mathrm{lep}}_{22}|$ &$2.471\times 10^{-9}$&$2.471\times 10^{-9}$   &$2.604\times 10^{-8}$&$2.606\times 10^{-8}$ \\
$|S^{\mathrm{lep}}_{23}|$ &$2.994\times 10^{-9}$&$2.994\times 10^{-9}$   &$3.378\times 10^{-8}$&$3.382\times 10^{-8}$ \\
$|S^{\mathrm{lep}}_{31}|$ &$7.810\times 10^{-9}$&$7.811\times 10^{-9}$   &$3.241\times 10^{-8}$&$3.244\times 10^{-8}$ \\
$|S^{\mathrm{lep}}_{32}|$ &$5.149\times 10^{-9}$&$5.150\times 10^{-9}$   &$2.137\times 10^{-8}$&$2.138\times 10^{-8}$ \\
$|S^{\mathrm{lep}}_{33}|$ &$1.413\times 10^{-9}$&$1.413\times 10^{-9}$   &$5.865\times 10^{-9}$&$5.866\times 10^{-9}$ \\ \hline
\end{tabular}
\end{center}
\end{table}

\begin{table}[t]
\caption{\label{Heavyparameter2} 
Possible ranges of $a_D, b_D, c_D, a_R, b_R, m_{1,2,3N}$ and $(\delta M_N)_{ij}$ (in GeV)}
\begin{center}
\begin{tabular}{l c c c c c c c c c c c c c |c|} \hline
\multirow{2}{0.75cm}{\centering } & \multicolumn{2}{c}{Normal odering} & \multicolumn{2}{c}{Inverted odering} \\
\cline{2-5}
& Minimum & Maximum  & Minimum  & Maximum  \\ \hline

$|a_{D}| $ &$1.740\times 10^{-3}$&$2.750\times 10^{-3}$  &$1.740\times 10^{-3}$&$2.750\times 10^{-3}$ \\
$|b_{D}| $ &$1.740\times 10^{-3}$&$2.750\times 10^{-3}$  &$1.740\times 10^{-3}$&$2.750\times 10^{-3}$ \\
$|c_{D}| $ &$1.740\times 10^{-3}$&$2.750\times 10^{-3}$  &$1.740\times 10^{-3}$&$2.750\times 10^{-3}$ \\ \hline

$|a_{R}| $ &$3.544\times 10^{5}$&$5.551\times 10^{5}$   &$3.372\times 10^{4}$&$5.266\times 10^{4}$ \\
$|b_{R}| $ &$1.652\times 10^{5}$&$2.579\times 10^{5}$   &$3.978\times 10^{4}$&$6.212\times 10^{4}$ \\ \hline

$m_{1N}$ &$1.650\times 10^{5}$&$2.584\times 10^{5}$     &$3.975\times 10^{4}$&$6.217\times 10^{4}$ \\
$m_{2N} $ &$2.543\times 10^{5}$&$5.556\times 10^{5}$    &$3.372\times 10^{4}$&$5.274\times 10^{4}$ \\
$m_{3N} $ &$1.650\times 10^{5}$&$2.584\times 10^{5}$    &$3.975\times 10^{4}$&$6.217\times 10^{4}$ \\ \hline

$|(\delta M_N)_{13}| $ &$1.465\times 10^{-11}$&$2.351\times 10^{-11}$  &$6.088\times 10^{-11}$&$9.758\times 10^{-11}$ \\
$|(\delta M_N)_{22}| $ &$5.449\times 10^{-12}$&$1.335\times 10^{-11}$  &$5.741\times 10^{-11}$&$1.403\times 10^{-10}$ \\
$|(\delta M_N)_{31}| $ &$1.465\times 10^{-11}$&$2.351\times 10^{-11}$  &$6.088\times 10^{-11}$&$9.758\times 10^{-11}$  \\ \hline

\end{tabular}
\end{center}
\end{table}

\newpage
\section{\label{conclusion}Conclusions}

We have constructed a SM extension based on $T^\prime\times Z_{10} \times Z_2$ symmetry 
for generating the expected neutrino mass matrix 
with the 
relation $(M_\nu)_{13}=(M_\nu)_{31}=-\frac{1}{2}(M_\nu)_{22}$ via the contributions of the Type-I seesaw mechanism and Weinberg-type operators. 
The proposed model possesses viable parameters capable of predicting the neutrino oscillation parameters in good agreement with recent experimental data.
Our analysis reveals the predicted regions for the physical quantities,  given as follows. The two mass squared splittings are $\delta m^2\in (69.360, 79.220)\, \mathrm{meV}^2$ and $\Delta m^2\in (2.484, 2.490)10^3\,\mathrm{meV}^2$ for normal ordering (NO) while $\delta m^2\in (69.450, 79.160)\, \mathrm{meV}^2$ and $\Delta m^2\in (-2.464, -2.456)10^3\,\mathrm{meV}^2$ for inverted ordering (IO).
The lightest neutrino mass is $m_{\ell}\in (36.720, 36.780)$ meV for NO and $m_{\ell}\in (62.220,\, 62.310)$ meV for IO. The sum of neutrino mass is $\sum m_\nu \in (136.700,\, 136.800)$ meV for NO and $\sum m_\nu \in (221.400,\, 221.600)$ meV for IO. Two Majorana phases are predicted to be $\alpha\in (6.367, 6.380)^\circ$ and $\beta\in (6.936, 6.946)^\circ$ for NO while $\alpha \simeq 358.800^\circ$ and $\beta \simeq 0.600^\circ$ for IO. Finally, the effective mass mass is $m_{\mathrm{ee}}\in (36.940, 36.980)$ meV for NO and $m_{\mathrm{ee}}\in (76.290, 76.360)$ meV for IO. Based on these results, the Yukawa-like couplings are estimated, which can  naturally explain the charged - lepton as well as neutrino mass hierarchies.

\section*{Acknowledgments}

This research is funded by Vietnam National Foundation for Science and Technology Development (NAFOSTED) under grant number 103.01-2023.45.

\newpage
\appendix

\section{\label{apptprime}$T^\prime$ group}
The non-Abelian discrete group $T^\prime$ is the double covering group of $A_4$ \cite{Kobayashi2022}. It has 24 elements divided into
seven conjugacy classes corresponding to seven reducible representations including three 
singlets 
$\underline{1}, \underline{1}^\prime, \underline{1}^{\prime\prime}$, three 
doublets 
$\underline{2}, \underline{2}^\prime, \underline{2}^{\prime\prime}$ and one
triplets 
$\underline{3}$.

Let us put $\underline{1}(a)$, $\underline{2}(a_1, a_2)$ and $\underline{3}(a_1, a_2, a_3)$ which respectively mean some
snglet, doublet and triplet representations of $T^\prime$ and similarly for the other ones. The
numbered multiplets 
$(...,a_i
b_j,...)$ where $a_i$ and $b_j$ denote the components of
some multiplet representations of $T^\prime$. In the basis where $p=i$ and $p_1=p_2=1$, the tensor product of all reducible representations of $T^\prime$  are determined as follows \cite{Kobayashi2022}: \\
$\bullet$ The tensor products of two singlets are given by
\begin{align}
&\underline{1} (a)\times \underline{1} (b)=\underline{1}^{\prime} (a)\times \underline{1}^{\prime\prime} (b)=\underline{1}^{\prime\prime} (a)\times \underline{1}^\prime (b)=\underline{1} (ab), \crn
&\underline{1}^\prime (a)\times \underline{1}^\prime (b)=\underline{1}^{\prime\prime} (ab),\hs \underline{1}^{\prime\prime} (a)\times \underline{1}^{\prime\prime} (b)=\underline{1}^\prime (ab).
\end{align}
$\bullet$ The tensor products of two doublets are given by
\begin{align}
&\left(
\begin{array}{ccc}
a_1\\
a_2\\
\end{array}%
\right)_{\underline{2} (\underline{2}^\prime)}\times \left(
\begin{array}{ccc}
b_1\\
b_2\\
\end{array}%
\right)_{\underline{2}(\underline{2}^{\prime\prime})}= \left(\frac{a_1b_2-a_2b_1}{\sqrt{2}}\right)_{\underline{1}}
+\left(
\begin{array}{ccc}
\frac{a_1b_2+a_2b_1}{\sqrt{2}}\\
-a_1b_1\\
a_2b_{2}\\
\end{array}%
\right)_{\underline{3}}, \\
&\left(
\begin{array}{ccc}
a_1\\
a_2\\
\end{array}%
\right)_{\underline{2}^\prime (\underline{2})}\times \left(
\begin{array}{ccc}
b_1\\
b_2\\
\end{array}%
\right)_{\underline{2}^\prime (\underline{2}^{\prime\prime})}=\left(\frac{a_1b_2-a_2b_1}{\sqrt{2}}\right)_{\underline{1}^{\prime\prime}}
+\left(
\begin{array}{ccc}
-a_1b_1\\
a_2b_{2}\\
\frac{a_1b_2+a_2b_1}{\sqrt{2}}\\
\end{array}%
\right)_{\underline{3}}, \\
&\left(
\begin{array}{ccc}
a_1\\
a_2\\
\end{array}%
\right)_{\underline{2}^{\prime\prime} (\underline{2})}\times \left(
\begin{array}{ccc}
b_1\\
b_2\\
\end{array}%
\right)_{\underline{2}^{\prime\prime} (\underline{2}^{\prime})}= \left(\frac{a_1b_2-a_2b_1}{\sqrt{2}}\right)_{\underline{1}^\prime}
+\left(
\begin{array}{ccc}
a_2b_{2}\\
\frac{a_1b_2+a_2b_1}{\sqrt{2}}\\
-a_1b_1\\
\end{array}%
\right)_{\underline{3}}.
\end{align}
$\bullet$ The tensor products of two triplets are given by
\bea
&&\left(a_1\hs
a_2\hs
a_3\hs
\right)_{\underline{3}}\times \left(
b_1\hs
b_2\hs
b_3\right)_{\underline{3}}=(a_1 b_1+a_2 b_3+a_3 b_2)_{\underline{1}}+(a_3 b_3+a_1 b_2+a_2 b_1)_{\underline{1}^\prime}\crn
&&\hspace{2.25 cm}+\,(a_2 b_2+ a_1 b_3+ a_3 b_1)_{\underline{1}^{\prime\prime}}+\left(a_2 b_3-a_3 b_2,
a_1 b_2-a_2 b_1,
a_3 b_1-a_1 b_3\right)_{\underline{3}_a}\crn
&&\hspace{2.25 cm} +\,\left(2 a_1 b_1-a_2 b_3-a_3 b_2,
2 a_3 b_3-a_1 b_2-a_2 b_1,
2 a_2 b_2-a_1 b_3-a_3 b_1\right)_{\underline{3}_s}.
\eea

\newpage
$\bullet$ The rules to conjugate the representations of $T^\prime$ are
\bea
&&\underline{1}^*(x^*)=\underline{1}(x^*),\hs
\underline{1}^{\prime *}(x^*)=\underline{1}^{\prime\prime}(x^*),\hs \underline{1}^{\prime\prime *}(x^*)=\underline{1}^\prime(x^*),
\hs \underline{3}^{*}\left(a^*_1,
a^*_2,
a^*_3
\right)=\underline{3}\left(a^*_1,
a^*_2,
a^*_3\right),\crn
&&\underline{2}^*\left(
\begin{array}{ccc}
a^*_1\\
a^*_2\\
\end{array}%
\right)=\underline{2}\left(
\begin{array}{ccc}
a^*_1\\
a^*_2\\
\end{array}%
\right), \hs \underline{2}^{\prime *}\left(
\begin{array}{ccc}
a^*_1\\
a^*_2\\
\end{array}%
\right)=\underline{2}^{\prime\prime}\left(
\begin{array}{ccc}
a^*_1\\
a^*_2\\
\end{array}%
\right), \hs \underline{2}^{\prime\prime *}\left(
\begin{array}{ccc}
a^*_1\\
a^*_2\\
\end{array}%
\right)=\underline{2}^{\prime}\left(
\begin{array}{ccc}
a^*_1\\
a^*_2\\
\end{array}%
\right). 
\eea
\section{\label{appz10}$Z_{10}$ group}
$Z_N$ group represents a symmetry that operates within the integers modulo N, it involves the numbers from 0 to $(N-1)$. The irreducible representation corresponding to an element of $Z_N$ group is characterized by an integer $n$, of the form $e^{i\left(\frac{2\pi}{N}\right)n}$ with
$n=0,1,...,(N-1)$. Therefore, the integer number $n$ is used to label the elements of the irreducible representations of $Z_N$ group. Suppose that $n_1$ and $n_2$ are two group elements of $Z_N$, they satisfy the 
relation
$n_1 \times  n_2 = (n_1 + n_2) \hspace{0.2 cm}\mbox{mod} \hspace{0.2 cm} N$. 
In the case where $N=10$, $Z_{10}$ group, the integer $n$ takes values from 0 to 9 and they satisfy, 
\begin{equation}
n_1 \times n_2 = (n_1+n_2) \hspace{0.2 cm}\mbox{mod} \hspace{0.2 cm}10.\label{Z10relation}
\end{equation}
\section{\label{preventedterm}Prevented terms}
All terms, up to dimension six, prevented by 
$T^\prime, Z_{10}$ and $Z_2$ are given in Table \ref{PreventedtermsT},
\begin{table}[ht]
\begin{center}
  \caption{\label{PreventedtermsT} Prevented terms}
 \begin{tabular}{|c|c|c|c|c|c|c|c|} \hline
Prevented terms $\big(\overline{\psi}_L l_{2R}=\overline{\psi}_L l_{3R};\, \, \overline{\psi}_L \nu_{1R}=\overline{\psi}_L \nu_{2R}=\overline{\psi}_L \nu_{3R}; $  & Prevented \\
$ \,\, \overline{\nu}^c_{1R} \nu_{1R}=\overline{\nu}^c_{2R} \nu_{3R}=\overline{\nu}^c_{3R} \nu_{2R}; \,\, \overline{\nu}^c_{2R} \nu_{2R}=\overline{\nu}^c_{1R} \nu_{3R}=\overline{\nu}^c_{3R} \nu_{1R}; \overline{\nu}^c_{3R} \nu_{3R}=\overline{\nu}^c_{1R} \nu_{2R}=\overline{\nu}^c_{2R} \nu_{1R} \big)$ & by \\  \hline

$(\overline{\nu}^c_{1R} \nu_{1R})\chi, (\overline{\nu}^c_{1R} \nu_{1R})\rho,
(\overline{\nu}^c_{1R} \nu_{1R}) (\Theta)_{\underline{1}}, (\overline{\nu}^c_{1R} \nu_{1R}) (\Theta)_{\underline{1}^{\prime\prime}}, (\overline{\nu}^c_{1R} \nu_{1R})\Omega;
(\overline{\nu}^c_{2R} \nu_{2R})\chi,
(\overline{\nu}^c_{2R} \nu_{2R})\rho,
$&\multirow{4}{0.35 cm}{$T^\prime$}\\

$(\overline{\nu}^c_{2R} \nu_{2R})\eta,   (\overline{\nu}^c_{2R} \nu_{2R})(\varphi^2 \eta)_{\underline{1}^\prime},
(\overline{\nu}^c_{2R} \nu_{2R})(\eta^2\eta^*)_{\underline{1}^\prime},
(\overline{\nu}^c_{2R} \nu_{2R}) (\Theta)_{\underline{1}^{\prime}},
(\overline{\nu}^c_{2R} \nu_{2R})(\Theta)_{\underline{1}^{\prime\prime}},
(\overline{\nu}^c_{2R} \nu_{2R}) \Omega;$&\\

$
 (\overline{\nu}^c_{3R} \nu_{3R})\chi, (\overline{\nu}^c_{3R} \nu_{3R})\rho,
(\overline{\nu}^c_{3R} \nu_{3R})_{\underline{1}^{\prime}}\eta,
(\overline{\nu}^c_{3R} \nu_{3R})(\varphi^2\eta)_{\underline{1}^\prime},
(\overline{\nu}^c_{3R} \nu_{3R}) (\eta^2\eta^*)_{\underline{1}^\prime},
(\overline{\nu}^c_{3R} \nu_{3R}) (\Theta)_{\underline{1}^{}}, $&\\
$(\overline{\nu}^c_{3R} \nu_{3R}) (\Theta)_{\underline{1}^{\prime}}, (\overline{\nu}^c_{3R} \nu_{3R})\Omega $&\\
\hline

$(\overline{\psi}_L l_{1R}) \Psi_1, \overline{\psi}_L l_{2R} \Psi_2;
(\overline{\psi}_L \psi^c_L)\Psi_3; (\overline{\psi}_L \nu_{iR})\Psi_4,
(\overline{\nu}^c_{1R} \nu_{1R}) \Omega_5, (\overline{\nu}^c_{2R} \nu_{2R}) \eta^3,
(\overline{\nu}^c_{2R} \nu_{2R}) \eta^{*3}, $&\multirow{2}{0.55 cm}{$Z_{10}$}\\

$(\overline{\nu}^c_{2R} \nu_{2R}) \Omega_5;
(\overline{\nu}^c_{3R} \nu_{3R}) \eta^*, (\overline{\nu}^c_{3R} \nu_{3R}) (\varphi^2\eta^*), (\overline{\nu}^c_{3R} \nu_{3R}) (\eta^{*2} \eta),
(\overline{\nu}^c_{3R} \nu_{3R}) (\eta^2\eta^*), (\overline{\nu}^c_{3R} \nu_{3R}) \Omega_5, $&\\ \hline

$(\overline{\psi}_L l_{1R})(H\chi), (\overline{\psi}_L l_{1R})(H\rho),
 (\overline{\psi}_L l_{2R})(H\varphi\chi), (\overline{\psi}_L l_{2R})(H\varphi\rho); (\overline{\psi}_L \nu_{iR})(\widetilde{H}\varphi \chi),
 (\overline{\psi}_L \nu_{iR})(\widetilde{H}\varphi \rho). $&$Z_2$\\ \hline

\end{tabular}
\end{center}
\end{table}

Here, we use the following notations,
\bea
\Theta &=&\phi^2\chi^*+
\phi^2\rho^*
+\phi \phi^*\chi
+\phi \phi^*\rho
+\phi^2\eta^*+\phi\phi^*\eta+\chi^2\chi^*+\chi^2\rho^*+\rho\rho^*\chi
+\chi\chi^*\rho \crn
&+&\chi^2\eta^* + \chi \rho\eta^*+\chi \chi^*\eta+\chi \rho^*\eta+\rho^2\chi^*+\eta^2\chi^*
+\chi^* \rho\eta+\rho\rho^* \eta+
\rho^2\eta^*,  \label{Theta}\\
\Omega&=&\varphi^2 \chi+\varphi^2 \rho+\eta\eta^*\chi+\eta\eta^*\rho+\eta^2\chi^*+\eta^2\rho^*, \label{Omega}\\
\Psi_1 &=&H\phi^*\varphi+H\varphi\chi+H\varphi\rho+H\varphi\chi^*+H\varphi\rho^*, \label{Psi1}\\
\Psi_2&=&H\phi^* +H\chi+H\rho+H\chi^*+H\rho^*+H\phi^2+H\phi^{*2}+H\phi\phi^{*}+H\chi^{2}+H\rho^{2} \crn
&+& H\chi^{}\rho + H\chi^{*2}+H\rho^{*2}
+H\chi^{*}\rho^*+H\chi^{}\rho^* +H\chi^{*}\rho+H\chi^{}\chi^*+H\rho^{}\rho^* \crn
&+& H\chi^{}\eta+H\rho^{}\eta +H\chi^{}\eta^*+H\rho^{}\eta^*
+H\chi^{*}\eta+H\rho^{*}\eta+H\chi^{*}\eta^*+H\rho^{*}\eta^*, \label{Psi2}\\
\Psi_3 &=&\widetilde{H}^2+\widetilde{H}^2\phi+\widetilde{H}^2\phi^*+\widetilde{H}^2\chi^*+ \widetilde{H}^2\rho^*+\widetilde{H}^2\eta^*,  \label{Psi3}\\
\Psi_4 &=&\widetilde{H}\phi^*+\widetilde{H}\chi+\widetilde{H}\chi^*+\widetilde{H}\rho+\widetilde{H}\rho^*+\widetilde{H}\phi^{2}
+\widetilde{H}\phi^{*2}+\widetilde{H}\phi \phi^{*}+\widetilde{H}\chi^{2}+\widetilde{H}\chi^{*2} +\widetilde{H}\chi\rho^{} \crn
&+&\widetilde{H}\rho^{2}
\widetilde{H}\rho^{*2}+\widetilde{H}\chi^*\rho^{*}+\widetilde{H}\chi\chi^{*}+\widetilde{H}\rho\rho^{*}
+\widetilde{H}\chi\rho^{*}+\widetilde{H}\chi^*\rho^{}+\widetilde{H}\phi\chi^{}
+\widetilde{H}\phi\chi^{*}+\widetilde{H}\phi\rho^{} \crn
&+&\widetilde{H}\phi\rho^{*}+\widetilde{H}\phi^*\chi^{}+\widetilde{H}\phi^*\rho^{}
+\widetilde{H}\phi^*\chi^{*}+\widetilde{H}\phi^*\rho^{*}
+\widetilde{H}\phi\eta^{} +\widetilde{H}\phi\eta^{*}+\widetilde{H}\phi^*\eta^{}
+\widetilde{H}\phi^*\eta^{*}
\crn
&+& \widetilde{H}\chi\eta^{}
+\widetilde{H}\rho\eta^{}
+\widetilde{H}\chi\eta^{*}
+\widetilde{H}\rho\eta^{*}
+\widetilde{H}\chi^*\eta^{}
+\widetilde{H}\rho^*\eta^{}
+\widetilde{H}\chi^*\eta^{*}
+\widetilde{H}\rho^*\eta^{*}, \label{Psi4}\\
\Psi_5&=&\phi^3+\phi^{*2}\phi+\phi^{2}\phi^*+\phi^{*3}
+\phi \chi^{2}+\phi \rho^{2}+\phi \chi \rho+\phi \chi^{*2}+\phi \rho^{*2}+\phi \chi^* \rho^*
+\phi \chi\chi^{*}\crn
&+&\phi \rho\rho^{*}+\phi \chi \rho^*+\phi \chi^* \rho
+\phi^2\chi+\phi^2\rho+\phi \phi^*\chi^*+\phi\phi^*\rho^*+\phi^2\eta
+\phi\phi^*\eta^*+\phi\chi\eta+\phi\rho\eta\crn
&+&\phi \chi\eta^*+\phi\rho\eta^*+\phi \chi^*\eta+\phi\rho^*\eta+\phi \chi^*\eta^*+\phi\rho^*\eta^*+\phi^*\chi^2+\phi^*\rho^2
+\phi^*\chi \rho
+\phi^*\chi^{*2}\crn
&+&\phi^*\rho^{*2}+\phi^*\chi^* \rho^*+\phi^*\chi\chi^{*}+\phi^*\rho\rho^{*}+\phi^*\chi \rho^*+\phi^*\chi^* \rho+\phi^{*2}\chi+\phi^{*2} \rho+\phi^{*2}\chi^*
+ \phi^{*2} \rho^*\crn
&+& \phi^{*2} \eta+\phi^{*2} \eta^*+\phi^*\rho \eta
+\phi^*\chi \eta^*+\phi^*\rho\eta^*+\phi^*\chi^* \eta+\phi^*\rho^*\eta
+\phi^*\chi^* \eta^*+\phi^*\rho^* \eta^* +\phi\phi^*\chi
\crn
&+& \chi^3
+\rho^2\chi
+\chi^2\rho
+\chi^{*2}\chi
+\rho^{*2}\chi
+\chi \chi^{*}\rho^*
+\chi \chi^{*}\eta
+\chi^2\eta
+\chi \rho\eta
+\chi \chi^{*}\eta^*
+\chi \rho^*\eta^*
\crn
&+&\chi^{*3} +\rho^{*2}\chi^*
+\chi^{*2}\rho^*
+ \chi^{*2}\chi
+\rho\rho^* \chi^{*}
+\chi^{*2}\rho
+\eta^{*2}\chi^*
+\eta \eta^*\chi^{*}
+\chi^{*}\rho^*\phi
+\chi^* \rho\eta^{*} \crn
&+&\chi^{*2}\eta
+\chi^* \rho^{*}\eta
+\chi^{*2}\eta^*
+\chi^* \rho^{*}\eta^*
+\phi \phi^{*}\eta
+\rho^{2}\eta
+\rho^{*2}\eta
+\rho^{*2}\eta^*
+\rho \rho^{*}\eta^*
+\phi^{*}\chi\eta.\label{Psi5}\eea
\newpage

\section{\label{potential}The minimization condition of the scalar potential}
The scalar potential 
is given by:
\begin{align}
V_{S}&= V(H)+V(\phi)+V(\varphi)+ V(\chi) + V(\rho) + V(\eta) +V(H, \phi)+V(H, \varphi) \crn
&+ V(H, \chi) + V(H, \rho)  + V(H, \eta) +V(\phi, \varphi)+ V(\phi, \chi) + V(\phi, \rho) + V(\phi, \eta)
  \crn
&+ V(\varphi, \chi) + V(\varphi, \rho)  + V(\varphi, \eta)
+ V(\chi, \rho)+ V(\chi, \eta)
+ V(\rho, \eta) , 
\label{Vtotal}
\end{align}
where
\bea
&&V(H)=\mu^2_H (H^{\+} H) +\lambda^H (H^{\+} H)^2, \hs V(\phi)=\mu^2_\phi (\phi^{*} \phi)_{\mathbf{1}} + \lambda^\phi \big[(\phi^{*} \phi)_{\mathbf{1}}^2
+(\phi^{*} \phi)_{\mathbf{1}^{'}} (\phi^{*} \phi)_{\mathbf{1}^{''}}\big], \label{Vphi} \nonumber \\
&&V(\varphi)=\mu^2_\varphi (\varphi^{*} \varphi) +\lambda^\varphi (\varphi^{*} \varphi)^2, \,
V(\chi)=V(\phi\rightarrow\chi), 
\, V(\rho)=V(\phi\rightarrow\rho), 
\, V(\eta)=V(\varphi\rightarrow\eta), 
\nonumber \\
&&V(H, \phi)=\lambda^{H \phi} (H^{\dagger} H)_{1}(\phi^*\phi)_{1}, \,
V(H, \varphi)=\lambda^{H \varphi} (H^{\dagger} H)_{1}(\varphi^*\varphi)_{1}, 
V(H, \chi)=V(H, \phi\rightarrow\chi), 
\crn
&&V(H, \rho)=V(H, \phi\rightarrow\rho), 
V(H, \eta)=V(H, \varphi\rightarrow \eta), 
V(\phi,\varphi)=\lambda^{\phi\varphi} \big[(\phi^{*} \phi)_{1}(\varphi^*\varphi)_{1}+(\phi^{*} \varphi)_{3}(\varphi^*\phi)_{3}\big], \hs  \label{Vphivarphi}\nonumber \\
&&V(\phi,\chi)=\lambda^{\phi\chi}_1 \big[(\phi^{*} \phi)_{1}(\chi^*\chi)_{1}+(\phi^{*} \chi)_{1}(\chi^*\phi)_{1}\big]
+\lambda^{\phi\chi}_2 \big[(\phi^{*} \phi)_{1^\prime}(\chi^*\chi)_{1^{\prime\prime}}+(\phi^{*} \chi)_{1^\prime}(\chi^*\phi)_{1^{\prime\prime}}\big] \crn
&&\hspace{1.35 cm}+\lambda^{\phi\chi}_3 \big[(\phi^{*} \phi)_{1^{\prime\prime}}(\chi^*\chi)_{1^{\prime}}+(\phi^{*} \chi)_{1^{\prime\prime}}(\chi^*\phi)_{1^{\prime}}\big]
+\lambda^{\phi\chi}_4 \big[(\phi^{*} \phi)_{3_s}(\chi^*\chi)_{3_s}+(\phi^{*} \chi)_{3_s}(\chi^*\phi)_{3_s}\big], \label{Vphichi} \nonumber \\
&&V(\phi,\rho)=V(\phi,\chi\rightarrow \rho), 
\hs V(\phi,\eta)=V(\phi,\varphi \rightarrow\eta), 
\hs V(\varphi,\chi)=V(\varphi, \phi\rightarrow\chi), 
\crn &&
V(\varphi,\rho)=V(\varphi,\chi\rightarrow\rho), 
\hs V(\varphi,\eta)=\lambda^{\varphi\eta} \big[(\varphi^{*} \varphi)_{1}(\eta^*\eta)_{1}+(\varphi^{*} \eta)_{1^\prime}(\eta^*\varphi)_{1^{\prime\prime}}\big], \nonumber \\
&&V(\chi,\rho)=V(\phi\rightarrow \chi,\rho), 
\hs V(\rho,\eta)=V(\chi\rightarrow\rho,\eta). 
\label{Vend}
\eea
For simplicity 
we consider the case where all the VEVs are real. We firstly consider the VEV alignments of $T^\prime$ triplets $\phi, \chi$ and $\rho$ whose general forms are $\langle\phi\rangle=(v_{\phi_1},\, v_{\phi_2},\, v_{\phi_3})$, $\langle\chi\rangle=(v_{\chi_1},\, v_{\chi_2},\, v_{\chi_3})$ and $\langle\rho\rangle=(v_{\rho_1},\, v_{\rho_2},\, v_{\rho_3})$, respectively. The minimization conditions on $v_{\phi_i}, v_{\chi_i}$ and $v_{\rho_i}\, (i=1,2,3)$ reads:
\bea
&&\lambda^\phi \left(2 v^3_{\phi_1}+8 v_{\phi_1} v_{\phi_2} v_{\phi_3}+v_{\phi_2}^3+v_{\phi_3}^3\right) + \mu_\phi^2 v_{\phi_1}  =0,\label{eqvphi1}\\
&&\lambda^\phi \left(3 v_{\phi_1} v_{\phi_2}^2+4 v_{\phi_1}^2 v_{\phi_3}+5 v_{\phi_2} v_{\phi_3}^2\right) + \mu_\phi^2 v_{\phi_3}=0,\label{eqvphi2}\\
&&\lambda^\phi  \left(3 v_{\phi_1} v_{\phi_3}^2+ 4 v_{\phi_1}^2 v_{\phi_2}+5 v_{\phi_2}^2 v_{\phi_3}\right)+\mu_\phi ^2 v_{\phi_2} =0,\label{eqvphi3}\\
&&\lambda^\chi  \left(2 v_{\chi_1}^3+8 v_{\chi_1} v_{\chi_2} v_{\chi_3}+v_{\chi_2}^3+v_{\chi_3}^3\right)+\mu_\chi ^2 v_{\chi_1} =0,\label{eqvchi1}\\
&&\lambda^\chi  \left(4 v_{\chi_1}^2 v_{\chi_3}+3 v_{\chi_1} v_{\chi_2}^2+5 v_{\chi_2} v_{\chi_3}^2\right)+\mu_\chi ^2 v_{\chi_3} =0,\label{eqvchi2}\\
&&\lambda^\chi  \left(4 v_{\chi_1}^2 v_{\chi_2}+3 v_{\chi_1} v_{\chi_3}^2+5 v_{\chi_2}^2 v_{\chi_3}\right)+\mu_\chi ^2 v_{\chi_2} =0,\label{eqvchi3}\\
&&\lambda^\rho  \left(2 v_{\rho_1}^3+8 v_{\rho_1} v_{\rho_2} v_{\rho_3}+v_{\rho_2}^3+v_{\rho_3}^3\right)+\mu_\rho ^2 v_{\rho_1} =0,\label{eqvrho1}\\
&&\lambda^\rho  \left(4 v_{\rho_1}^2 v_{\rho_3}+3 v_{\rho_1} v_{\rho_2}^2+5 v_{\rho_2} v_{\rho_3}^2\right)+\mu_\rho ^2 v_{\rho_3} =0,\label{eqvrho2}\\
&&\lambda^\rho  \left(4 v_{\rho_1}^2 v_{\rho_2}+3 v_{\rho_1} v_{\rho_3}^2+5 v_{\rho_2}^2 v_{\rho_3}\right)+\mu_\rho ^2 v_{\rho_2} =0. \label{eqvrho3}
\eea
 One of the solutions of the system of Eqs. (\ref{eqvphi1})-(\ref{eqvrho3}) 
 takes the following form:
\bea
&&v_{\phi_1}=
\frac{|\mu_\phi|}{\sqrt{2 \lambda^\phi }}\equiv v_{\phi}, \hs v_{\phi_2}=v_{\phi_3}=0, \hs
v_{\chi_1}=
\frac{|\mu_\chi|}{\sqrt{2 \lambda^\chi }}\equiv v_{\chi}, \hs v_{\chi_2}=v_{\chi_3}=0, \\
&&v_{\rho_1}=0, \hs v_{\rho_2}= 
\frac{|\mu_\rho|}{\sqrt{5 |\lambda^\rho|}} \equiv v_{\rho}=-v_{\rho_3},
\label{vphisolution}
\eea
which are exactly the same as 
those of $\phi, \chi$ and $\rho$ in Eq. (\ref{VEValig}).

Next, we prove that the scalar VEVs in Eq. (\ref{VEValig}) satisfy the
minimization condition of $V_{S}$ in Eqs. (\ref{Vtotal})-(\ref{Vend}).
Indeed, the  minimum condition of $V_{S}$, $\frac{\partial V_{S}}{\partial v_{\Psi}} =0$ and $\frac{\partial^2 V_S}{\partial v_{\Psi}^2}>0\,\, \left(v_\Psi=\{v, v_\phi, v_\varphi, v_\chi, v_\rho, v_\eta\}\right)$ provide the following solutions and stability conditions,
\bea
&&\mu^2_H=-2 \lambda^{H} v^2-\lambda^{H\eta} v_{\eta}^2+2 \lambda^{H\rho} v_{\rho}^2-\lambda^{H\varphi} v_{\varphi}^2-\lambda^{H\chi} v_{\chi}^2-\lambda^{H\phi} v_{\phi}^2,\label{muHsq}\\
&&\mu^2_\phi=-\lambda^{H\phi} v^2-2 \left(\lambda^{\phi \eta}  v_{\eta}^2+\lambda^{\phi \varphi}  v_{\varphi}^2+\lambda^{\phi} v_{\phi}^2\right)+ v_{\rho}^2 \big(2 \lambda^{\phi \rho}_1+\lambda^{\phi \rho}_2+\lambda^{\phi \rho}_3-2 \lambda^{\phi \rho}_4\big)\crn
&&\hspace{0.925 cm} -2 v_{\chi }^2 (\lambda^{\phi \chi}_1+4 \lambda^{\phi \chi}_4), \label{muphisq}\\
&&\mu^2_\varphi= - \lambda^{H\varphi} v^2-2 \left(\lambda^{\varphi \eta}  v_{\eta}^2-2 \lambda^{\varphi \rho}  v_{\rho}^2+\lambda^\varphi  v_{\varphi}^2+\lambda^{\varphi \chi}  v_{\chi}^2+\lambda^{\phi \varphi}  v_{\phi}^2\right), \label{muvarsq}\\
&&\mu^2_\chi=-\lambda^{H\chi} v^2-2 \left(\lambda^{\chi \eta}  v_{\eta}^2+\lambda^{\varphi \chi}  v_{\varphi}^2+\lambda_{\chi}  v_{\chi}^2\right)+v_{\rho}^2 (2 \lambda^{\chi \rho}_1+\lambda^{\chi \rho}_2+\lambda^{\chi \rho}_3-2 \lambda^{\chi \rho}_4)\crn
&&\hspace{0.925 cm} -2 v_{\phi}^2 (\lambda^{\phi \chi}_1+4 \lambda^{\phi \chi}_4), \label{muchisq}\\
&&\mu^2_\rho=-\lambda^{H\rho} v^2-2 \lambda^{\rho \eta}  v_{\eta}^2+5 \lambda^\rho  v_{\rho}^2-2 \lambda^{\varphi \rho} v_{\varphi}^2-\frac{v_\chi^2}{2}  \left(2 \lambda^{\chi \rho}_1+\lambda^{\chi \rho}_2+\lambda^{\chi \rho}_3-2 \lambda^{\chi \rho}_4\right)\crn
&&\hspace{0.925 cm} -\,\frac{v_\phi^2}{2} \big(2 \lambda^{\phi \rho}_1+\lambda^{\phi \rho}_2+\lambda^{\phi \rho}_3-2 \lambda^{\phi \rho}_4\big), \label{murhosq}\\
&&\mu^2_\eta= -\lambda^{H\eta} v^2-2 \left(\lambda^{\eta} v_{\eta}^2-2 \lambda^{\rho \eta}  v_{\rho}^2+\lambda^{ \varphi \eta}  v_{\varphi}^2+\lambda^{\chi \eta} v_{\chi}^2+\lambda^{\phi \eta} v_{\phi}^2\right),  \label{muetasq}
\eea
and
\bea
 \lambda^{H}  >0, \hs \lambda^{\phi} >0, \hs
\lambda_\varphi>0, \hs
\lambda_\chi>0, \hs
\lambda_\rho>0, \hs
\lambda_\eta>0.\label{ineq}&&
\eea


\end{document}